\documentstyle [12pt]{article}
\def\fnote#1#2{\begingroup\def\thefootnote{#1}\footnote{#2}\addtocounter
{footnote}{-1}\endgroup}
\evensidemargin=-.25in \textwidth=6.7in \def\half{{\textstyle{1\over2}}}
\def\pr#1{{\sl Phys.~Rev.~\bf D#1}}
\def\cpc#1{{\sl Comp.~Phys. Comm.~\bf #1}}\def\prz#1{{\sl Phys.~Rev.~\bf#1}}
\def\anp#1{{\sl Ann.~Phys.~(NY) \bf #1}}
\def\rmp#1{{\sl Rev.~Mod.~Phys. \bf #1}}
\def\etal{{\it et al.}}\def\sqstev{$\sqrt s=40$~TeV}
\def\etatwet{$|\eta|<2.8$}

\def\sixt{{\textstyle{1\over6}}}
\def\VEV#1{\langle #1 \rangle}
\def\sqstev{$\sqrt s=40$~TeV}
\begin{document}
\hfill UTHEP--92--0901 \vskip.01truein \hfill{June 1992}\vskip1truein
\centerline{\Large Radiative corrections in processes at the SSC
\fnote{\ast}{Supported in part by the Texas National Research Laboratory
Commission for the Superconducting Super Collider Laboratory via grant
RCFY9201, by the US Department of Energy Contracts DE-FG05-91ER40627
and DE-AC03-76ER00515, and by the Polish Ministry of Education grants
KBN 223729102 and KBN 203809101.}}
\vskip.25truein
\begin{center}
           {\sc G.~Siopsis,
\fnote{\dagger}{Talk presented at the XXXII Krak\'ow School of Theoretical
      Physics, Zakopane, June 1992.}
           D.~B. DeLaney\\}
           {\it Department of Physics and Astronomy\\
          The  University of Tennessee, Knoxville, TN 37996--1200, USA\\}
           {\sc S.~Jadach \\}
{\it Institute of Nuclear Physics, ul. Kawiory 26a, Krak\'ow, Poland\\
                  and\\
                Theory Division, CERN, Geneva 23, Switzerland \\}
          {\sc Ch.~Shio\\}
          {\it Department of Physics and Astronomy\\
           The University of Tennessee, Knoxville, TN 37996--1200, USA\\}
          {\sc B.~F.~L.~Ward \\}
          {\it Department of Physics and Astronomy\\
           The University of Tennessee, Knoxville, TN 37996--1200, USA\\
                and\\
     SLAC, Stanford University, Stanford, CA 94309, USA\\}
\end{center}
\vskip.05truein \baselineskip=21pt
\vskip.25truein\centerline{\bf ABSTRACT}\vskip.1truein

We discuss radiative corrections for interactions in the SSC environment.
Based on the theory of Yennie, Frautschi and Suura,
we develop appropriate Monte Carlo event generators to compute the
background electromagnetic radiation.
Our results indicate that multiple-photon effects must be taken into
account in the study of SSC physics such as Higgs decay.

\renewcommand\thepage{}\vfill\eject
\parskip.1truein \parindent=20pt \pagenumbering{arabic}
\section{Introduction}

The Superconducting Super Collider (SSC), which is scheduled to begin
operation in a few years, will probe a new frontier in high energy
physics.
New physical processes are expected to be discovered.
We then need to determine whether these processes fall within our
present understanding of particle interactions (Standard Model);
otherwise a new theory will have to be developed.
Thus, before the SSC is constructed,
we ought to extract predictions as precise as possible from our current
theory,
in order to maximize the discrimination between signal and background.
This amounts to calculating the higher-order radiative corrections on
SSC processes.

We have embarked on a calculation of the multiple-photon(gluon)
radiative effects in the SSC environment.
Here, we report on the progress we have made to date.
The basic tool in our study is a Monte Carlo event generator
algorithm which was
developed by two of us (S.J. and B.F.L.W.~\cite{cjw}) for high precision
$Z^0$ physics at LEP/SLC.
The algorithm relies on the theory of Yennie, Frautschi and Suura
\cite{yfs},
who have obtained expressions for scattering cross-sections that are
explicitly free from infrared divergences to all orders in the
electromagnetic coupling constant.
We utilize these expressions to calculate scattering amplitudes for SSC
processes with multiple photon production.
So far we have concentrated on initial-state electromagnetic radiation
\cite{djs}.
For a complete study of radiative effects, one needs to include final-state
radiation as well as the production of gluons.
We shall report on progress in this direction soon \cite{ddl}.

We consider the scattering of two fermions,
$f_1f_2 \to f_1'f_2'+n\gamma$,
where $n$ photons are emitted from the initial fermions, $f_1$ and $f_2$.
Our discussion is organized as follows.
In section 2 we review the YFS theory. In section 3 we discuss how the YFS
expressions can be combined with Monte Carlo methods to develop an event
generator (YFS2) that calculates multiple photon emission with arbitrary
detector cuts~\cite{cjw}.
In section 4 we extend YFS2 to the SSC domain, arriving at the event generator
SSCYFS2~\cite{djs}.
In section 5 we present sample results and comment on their
significance.
Finally in section 6 we discuss our conclusions.

\section{The Yennie-Frautschi-Suura theory}

Even though Green functions in quantum electrodynamics are not in
general infrared finite,
as shown by Yennie, Frautschi and Suura \cite{yfs},
all cross-sections that can be experimentally
observed are finite to all orders in perturbation theory.
The divergences arising from the infrared region of loop diagrams are
canceled by the effects of the undetectable soft photons.
To understand how this occurs, consider scattering of two fermions,
\begin{equation}
f_1(p_1)+f_2(p_2)\longrightarrow f_1'(q_1)+f_2'(q_2) \,\;,\label{gsp}
\end{equation}
where $p_1^\mu,p_2^\mu$ ($q_1^\mu,q_2^\mu$) are the momenta of the
incoming (outgoing) fermions, respectively.
To lowest order in the QED coupling constant $\alpha$, the cross-section is
the Born cross-section $d\sigma_B$.
To first order in $\alpha$, the cross-section can be written
(we only consider initial-state renormalization)
$d\sigma_B(1+2\alpha {\rm Re}B)$,
where
\begin{equation}
B(p_1,p_2)={i\over8\pi^2}\int {d^4k\over k^2}\left({2p_1-k\over k^2
-2p_1\cdot k}-{2p_2+k\over k^2+2p_2\cdot k}\right)^2\label{xti}
\end{equation}
is infrared divergent.

Now consider the same process where a soft photon is emitted from one of the
incoming fermions (bremsstrahlung).
By integrating over the momentum $k^\mu$ of the photon,
where $k^0 < \epsilon \sqrt s /2$, $s=(p_1+p_2)^2$ being the invariant
squared mass of the incoming state,
we obtain the cross-section
$d\sigma_B(1+\int {d^3 k \over k^0}\widetilde S)$,
where
\begin{equation}
\widetilde{S}(p_1,p_2,k)=-{\alpha\over4\pi^2}\left({p_1\over p_1\cdot
k}-{p_2\over p_2\cdot k}\right)^2\label{sti}
\end{equation}
is infrared divergent.
The parameter $\epsilon$ that separates hard from soft photons is
arbitrary and is chosen appropriately to match detector capabilities.

It can easily be seen that the divergences of $B$ and
$\widetilde S$ cancel each other.
Thus the experimentally observable cross-section
\begin{equation}
d\sigma = d\sigma_B\left( 1+2\alpha {\rm Re} B(p_1,p_2)
+ \int {d^3 k\over k^0} \widetilde S (p_1,p_2,k) \right)\label{gti}
\end{equation}
is infrared finite.

As was shown in~\cite{yfs}, this cancellation of infrared
divergences occurs to all orders in $\alpha$.
The resulting expression for the cross-section is manifestly infrared
finite.
It can be written in the form
$d\sigma = \exp\{\delta_{YFS}\} d\sigma_0$, where
\begin{eqnarray}
\begin{array}{rcl}
\delta_{YFS}(p_1,p_2,\epsilon) &=& 2\alpha{\rm Re}B+\int{d^3k\over
k^0}\widetilde{S}(p_1,p_2,k)\,(1-\theta(k^0-\epsilon\sqrt s/2))\\
&=& {\alpha\over\pi}\left(2(\ln(s/m_1m_2)-1)\ln\epsilon
+\half\ln(s/m_1m_2)-1+\pi^2/3\right)\,\;,
\end{array}
\label{fyf}
\end{eqnarray}
$m_1$, $m_2$ being the masses of the incoming fermions.
The exponential form factor $\exp\{\delta_{YFS}\}$
is the soft-photon contribution summed to all
orders in perturbation theory.
We can implement such expressions in event generators,
in order to calculate the effects of radiation to arbitrary precision.

\section{The Monte Carlo generator YFS2 Fortran}

In this section, we discuss how one can utilize the exponentiation of
soft photon effects to develop Monte Carlo event generators that
estimate the effects of radiation.
Specifically, we review the key ingredients in
the Monte Carlo event generator YFS2 Fortran \cite{cjw} that was developed for
$e^+e^-\to f\bar{f}+n\gamma$, $f\neq e$, in the $Z^0$ energy regime.

We wish to use Monte Carlo methods to calculate the cross-section of the
process
\begin{equation}
e^+(p_1)+e^-(p_2)\longrightarrow f(q_1)+\bar{f}(q_2)+
\gamma(k_1)+\dots+\gamma(k_n)\,\;,\label{bsp}
\end{equation}
where we sum over the number of photons and integrate over their
momenta. Thus we include both soft and hard photons.
As outlined in the previous section, the differential
cross-section takes the form
\begin{equation}
d\sigma = \exp\{\delta_{YFS}\} \sum_{n=0}^\infty d\sigma^{(n)}\;\,,
\label{gsn}
\end{equation}
where $d\sigma^{(n)}$ is the contribution of $n$ hard photons.
It can be expressed in terms of the YFS hard-photon
residuals~\cite{yfs} $\bar\beta_i(k_1,\dots,k_i)$
($i=1,\dots,n$),
which are free of all virtual and real infrared divergences to all
orders in the QED coupling constant $\alpha$.
We obtain
\begin{eqnarray}
\begin{array}{rcl}
d\sigma^{(n)} &=& \left( \widetilde{S}(k_1)\cdots\widetilde{S}(k_n)
\overline\beta_0+\dots+\overline\beta_n(k_1,\dots,k_n)\right)
\\
&\times &{1\over n!}\delta^4\left(p_1+p_2-q_1-q_2-\sum_{i=1}^n k_i\right)
{d^3q_1\over q_1^0}{d^3q_2\over q_2^0}{d^3k_1\over k_1^0}\cdots
{d^3k_n\over k_n^0}\,.
\end{array}
\label{dsi}
\end{eqnarray}
We shall discuss these residuals shortly.

Events are generated as follows.
First the complicated cross-section $d\sigma$ is replaced by $d\sigma'$,
so that the integral $\int d\sigma'$ is simple enough to calculate.
We should emphasize that, for efficient event generation, it is always
desirable to generate a background population of events according to a
distribution which embodies all of the general features of
Eq.(\ref{dsi}), but which removes unnecessary details.
In addition, several changes of variables are used to make the background
generation simpler and more efficient from the standpoint of CPU time.

The variables in $d\sigma'$ are then generated according to this
distribution.
For each set of values (event) we calculate the weight
\begin{equation}
w={d\sigma \over d\sigma'}\;\,,\label{gwe}
\end{equation}
and then reject events according to their weights (\ref{gwe}).
The exact total cross-section is
\begin{equation}
\sigma=\int d\sigma' \langle w \rangle\;\,,\label{gts}
\end{equation}
where $\langle w \rangle$ is the average weight.
This procedure is very complicated due to the large number of variables
in $d\sigma'$.
In fact, even the number of variables, which is the dimension of phase
space, is not fixed and needs to be generated.

After several simplifications \cite{cjw} we arrive at the result
\begin{equation}
{\sigma'}^{(n)} (s)=\int_\epsilon^{v_{max}} {dv\over v} J_0(v)
{1\over (n-1)!}\left( {2\alpha\over\pi} \ln (s/m_e^2)
\ln (v/\epsilon) \right)^{n-1} \sigma_B (s)\;\,,\label{gsv}
\end{equation}
where $J_0$ is a certain Jacobian,
and $v=1-s'/s$, $s'=(q_1+q_2)^2$ being the squared mass of the
outgoing state, measures the fraction of energy
carried away by radiation.
$\sigma_B (s)$ is the total Born cross-section.
Thus we see that the number of hard photons ought to be generated
according to a {\it Poisson} distribution with mean
\begin{equation}
\overline n = {2\alpha\over\pi} \ln (s/m_e^2)
\ln (v/\epsilon ) \;\,.\label{gpv}
\end{equation}
The variable $v$ is generated according to the above integral (\ref{gsv})
with the help of a one-dimensional Monte Carlo generator.
Once $v$ and $n$ are determined, the phase-space variables $q_1,q_2,k_1,
\dots,k_n$ are generated analytically.
It is then straightforward, albeit cumbersome,
to compute the event weight (\ref{gwe}).

We now turn to a discussion of the hard-photon
residuals~\cite{yfs} $\bar\beta_i(k_1,\dots,k_i)$.
They contain the physics besides QED effects and are therefore model
dependent.
For YFS2, only the residuals $\overline
\beta_{0,1,2}$ are used, as two of us have explained in Ref.~\cite{cjw}.
We should emphasize that $\overline\beta_2$ has only been included in
the second-order leading-log approximation because we have
checked~\cite{sjb} that the exact result does not
affect the results of the program beyond the quoted $0.1\%$ accuracy
of the program.
The residuals are linear combinations of Born cross-sections whose
arguments are defined in a {\it reduced} phase space.
This means that in the YFS2 Monte Carlo, some choice must be made
for the reduction of the $n$-photon $+f\bar f$ phase space to the $j$-photon
$+f\bar f$ phase space ($n=0,1,2,\dots$, $j=0,1,2$, $n\geq j$), which is
involved in the definition of the residuals $\overline\beta_i$ ($i=0,1,2$).
We call this choice the reduction procedure ${\cal R}$ and the exact YFS result
(\ref{gsn}) is independent of it, if it is done according to the
rigorous YFS theory. The map ${\cal R}$ is such that momentum is
conserved in the reduced phase space.
It cannot depend on the individual
photon momenta and has to reduce to the identity in the limit of
vanishing photon momenta, but it is otherwise arbitrary.
This freedom can be exploited to minimize the weights and therefore the
error for a given number of simulated events.
Our choice for ${\cal R}$ is explained in~\cite{cjw}.

In this way we have realized the YFS theory for $e^+e^-\to f\bar f
+n\gamma$ with $\overline\beta_0$, $\overline\beta_1$, and $\overline\beta_2$.
Next, we discuss how we extend YFS2 to more general processes,
as well as the modifications
needed to make the program applicable in the SSC energy regime.

\section{Going from YFS2 to SSCYFS2}

An extension of the YFS2 Monte Carlo algorithm described in the previous
section to particle interactions at SSC energies
involves the introduction of new physics (mainly through a modification of
the $\overline \beta_i$, which we will currently achieve via the
Born cross-section), numerical problems (due to the very high energies
involved, care is needed for the accuracy of the formulas), and certain
technical problems associated with the Monte Carlo weight rejection method.

We begin by discussing the modifications made to introduce the new
physics at SSC energies. The new program, SSCYFS2,
computes the cross-section for the scattering of two fermions
\fnote{\natural}{At present, the program cannot handle third-generation
fermions.}
\begin{equation}
f_1(p_1)+f_2(p_2)\longrightarrow f_1'(q_1)+f_2'(q_2)
+\gamma(k_1)+\dots+\gamma(k_n)\,\;.\label{ffn}
\end{equation}
The mass parameters $m_q$ used for the quarks are the Lagrangian quark masses.
We have in mind that the overall momentum transfer in the interactions will be
large compared to the typical momenta inside the proton. In fact strictly
speaking, these quark
mass parameters should be running masses $m_q(\mu)$, where
$\mu$ is the scale at which they are being probed. Such a running mass effect
is well-known and is readily incorporated in the program, as the accuracy one
is interested in may dictate. Thus, with this understanding, further explicit
reference to the running mass effect is suppressed.

The interactions realized by YFS2~(Eq.(\ref{bsp})) involve only an exchange of
$\gamma$ and $Z^0$ in the $s$-channel.
Since SSCYFS2 realizes the more general interaction~(\ref{ffn}),
one needs to introduce $\gamma$, $Z^0$ and $W^\pm$ exchange in the $t$- and
$u$-channels, accordingly. This is done by generalizing
the Born cross-section to include the additional channels. Moreover, in the
case of quark interactions, a gluon exchange was added in all three channels.
The running strong coupling constant
\begin{equation}
\alpha_s(\mu)={12\pi\over(33-2n_f)\ln\{\mu^2/(\Lambda_{n_f}^ %check \{\} sizes
{\overline{MS}})^2\}}\,\;,\label{sal}
\end{equation}
was used, where $n_f$ is the number of
quark flavors below the energy level $\mu$. In our case, $n_f=6$, and therefore
the QCD parameter $\Lambda_6^{\overline{MS}}$ is used. It can easily be related
to the experimentally measured parameter $\Lambda_4^{\overline{MS}}=238\,MeV$:
\begin{equation}
\Lambda_6^{\overline{MS}}=\Lambda_5^{\overline{MS}}\left({\Lambda_5
^{\overline{MS}}\over m_t}\right)^{\!2/21},\,\;
\Lambda_5^{\overline{MS}}=\Lambda_4^{\overline{MS}}\left({\Lambda_4
^{\overline{MS}}\over m_b}\right)^{\!2/23}.\label{lms}
\end{equation}
The masses of the top and bottom quarks were set to $m_b=5$~GeV and
$m_t=250$~GeV, respectively, but the results
are little affected by their precise values.

Certain numerical problems arise at very high energies, because of the very
small value of all ratios $m/\sqrt s$, where $m$ is the mass of any interacting
particle, and \sqstev\ is the energy of the incoming fermions in
their center-of-mass frame.
Certain formulas had to be rewritten so that such small numbers would not be
ignored by the computer when they should not be; if one is not careful, ratios
of the form $0/0$ appear at various places. Working at SSC energies, however,
has the advantage that all terms of order $m^2/s$ or higher can be dropped.
The error is negligible and leads to a considerable simplification of formulas,
and consequently to a reduction in computer time.

Next we discuss the event-generation procedure.
In simplifying the exact cross-section $d\sigma$ in YFS2,
the residuals $\overline\beta_i$ ($i=1,2$) where set to zero,
whereas $\overline\beta_0$ was replaced by a constant.
In the present case, this is no longer possible, because of the presence of the
$t$-channel. The cross-section has a singularity at $t\equiv(p_1-q_1)^2=0$ of
the form $1/t^2$. To account for the singularity, an angle cutoff
$\theta_0=100$~mrad is introduced. This is in accord with current detector
capabilities, and can be changed at will. In the crude cross-section
$d\sigma'$ the residuals $\overline\beta_i$ ($i=1,2$) are still set to zero,
but $\overline\beta_0$ takes the form
\begin{equation}
A+{B\over t^2}\;,\label{scr}
\end{equation}
where the constants $A$ and $B$ depend on the interaction.
\fnote{\ddagger}{A fictitious photon mass cutoff was also tested,
but it turned out to lead to a large weight rejection rate.}
When a $u$-channel also contributes, a similar term of the form $1/u^2$ must
be added to account for the singularity at $u\equiv(p_1-q_2)^2=0$.

Finally, we comment on the choice of the reduction procedure which is needed
for the definition of the arguments of the YFS residuals $\overline\beta_i$
($i=0,1,2$), as explained in section~3. The reduction procedure is more
delicate in the presence of the $t$-channel, due to the singularity at $t=0$.
One has to make sure that the weights (\ref{gwe})
do not become uncontrollably large.
This is managed by making $t$ as large as possible after the reduction.
It is not always possible to increase the reduced $t$ so that the
weight (\ref{gwe})
remains below the maximum weight. The object of this exercise is to minimize
the error originating from the tail of the distribution of weights above the
maximum weight, which can be changed at will.
This is accomplished by a somewhat involved reduction
procedure, which is an adaptation of the similar procedure in
BHLUMI1.xx~\cite{jwb}.

This concludes our discussion of the modifications in the YFS2 program
which are necessary in order to realize interactions at SSC energies.

\section{Results}

We now present some results on the effects of multiple-photon
initial-state radiation on the incoming $qq$ and $q\bar q$ ``beams'' at
SSC energies using our YFS Monte Carlo event generator SSCYFS2 Fortran.
Our objective is to determine the size of these effects with an eye toward
their incorporation into SSC physics event generators.
For definiteness, we will illustrate our results with $q=u,d$, where we use
$m_u=5.1\times10^{-6}$~TeV, $m_d=8.9\times10^{-6}$~TeV, and view \sqstev\ as
our worst-case scenario. The results are similar in the more
typical~\cite{eei} case of center-of-mass energy $\sqrt
s\approx\sixt40$~TeV $\approx6.7$~TeV.
For these respective input scenarios, we shall discuss the following
distributions: the number of photons per event, the value of $v=(s-s')/s$,
and the squared transverse momentum of the outgoing
$n\gamma$ state.  These distributions give us a view of the effect of this
multiple-photon radiation on the incoming quarks and in the SSC
environment, where one is really interested in $p p\to H+X$,
where $H$ is the Standard Model Higgs particle.

Considering first the number of photons per event, we have the results in
Fig.~1. There, we show that for the $uu$ incoming beams,
the mean number $\VEV{n_\gamma}$ of radiated photons is $0.85\pm0.92$.
This should be compared to the $dd$
incoming state, where $\VEV{n_\gamma}=0.21\pm0.45$.
For reference, we recall~\cite{cjw} that at LEP/SLC energies,
the corresponding value of $\VEV{n_\gamma}$ is, for the incoming
$e^+e^-$ state, $\sim1.5\pm1.0$. Hence, we see here one immediate effect of the
high energy of the SSC incoming beams: the initial $uu$-type state will radiate
a significant number of real photons, with a consequent change in the observed
final-state character. In particular, the issue of how much energy is lost to
photon radiation is of immediate interest.
This energy is unavailable for Higgs production by $uu$ (or $dd$)
and, further, it may fake a signal of $H\to\gamma\gamma$ if we are unlucky.
Accordingly, we now look at the predicted distribution of the variable $v$.
If only one photon is radiated,
$v$ is just the energy of this photon in the center-of-mass system of the
incoming beams (in units of the incoming beam energy).

What we find for $v$ is shown in Fig.~2 for the $uu\to uu+n\gamma$ case
(the $dd\to dd+n\gamma$ case is similar). We see the expected shape of $v$ from
Ref.~\cite{cjw}, and its average value is $\VEV{v}=0.05\pm0.09$.
Hence, $\sim10\%$ of the incoming energy is radiated into photons;
this energy is not available for Higgs production and it is therefore crucial
to
fold our radiation into the currently available SSC Higgs production Monte
Carlo event generators~\cite{fep} and to complete the development of our own
YFS multiple-photon (-gluon) Higgs production Monte Carlo event generator,
which is under development.

Given that we know that in the SSC environment we have
significant multiple-photon
radiation effects, the question of immediate interest is how often the
transverse momenta of two photons are large enough that they could fake a
$H\to\gamma\gamma$ signal. We will answer this very important question in
detail in the not-too-distant future when our complete Higgs production YFS
Monte Carlo event generators are available~\cite{ddl}.
However, here we can begin
to study this question by looking into the transverse momentum distribution of
our YFS multiple-photon radiation in, e.g., $uu\to uu+n\gamma$.
This is shown in Fig.~3, where we plot the total transverse
momentum distribution of the respective YFS multiple-photon radiation.
What we find is that, for $\sqrt s = 40$~TeV,
the average value of this total transverse momentum is
(in the incoming $uu$ center-of-mass system)
\begin{equation}
\VEV{p_{\perp,tot}}\equiv\VEV{|\sum_{i=1}^n\vec{k_{i\perp}}|}
=(0.0184\pm0.0129)\sqrt s\,\;,\label{pper}
\end{equation}
where $k_i$ ($i=1,\dots,n$) are the four-momenta of the $n$ photons.
Hence, for the SDC acceptance cut of $\half|\ln\tan(\theta/2)|\equiv
|\eta|<2.8$, or $\theta_i>122$~mrad, this means that there may be some possible
background to $H\to\gamma\gamma$ for, e.g., $m_H\approx150$~GeV.

\section{Conclusion}

To summarize, our initial study of multiple-photon radiation in the SSC
physics environment shows that any Monte Carlo event generator which hopes to
achieve an accuracy of order~$10\%$ in the SSC physics simulations must treat
the respective effects in a complete way.
We have computed these effects for incoming quark-(anti)quark
states at SSC energies using the Monte Carlo event generator SSCYFS2 Fortran
based on our original YFS2 Monte Carlo in Ref.~\cite{cjw}.
We found that for an initial $uu$ state,
the mean number of radiated photons is $0.85\pm0.92$, so that the
multiple-photon character of the events must be taken into account in detailed
detector simulation and physics analysis studies. Furthermore,
the mean value of
$v=(s-s')/s$ is $0.05\pm0.09$ and the average total squared transverse momentum
$\VEV{k_{\perp,tot}^2}$ is $(0.025\pm0.002)$~s.
Hence, the impact of these event characteristics on Higgs production in general
and on the $H\to\gamma\gamma$ scenario in particular must be assessed in
detail.

At the moment, we can say that the initial platform for precision SSC
electroweak physics simulations on an event-by-event basis using our YFS
Monte Carlo approach~\cite{cjw} has been established.
A lot of work remains to be done. We need to assess the effects of gluon
radiation.
We also have to include final-state radiation and more physics.
This will allow us to perform a detailed study of processes of interest,
such as Higgs production.
On the technical side, a more efficient algorithm is needed that
utilizes a less cumbersome reduction procedure.
We enthusiastically  look forward to the complete development of our program.

\noindent{\large \ {ACKNOWLEDGMENTS:}}

One of the authors (G.S.) thanks the Organizers of the XXXII Krak\'ow
School of Theoretical Physics for inviting him to participate in their
conference. In addition, two of the authors (S.J. and B.F.L.W.)
thank Profs. C.~Prescott and J.~Ellis of SLAC and CERN respectively
for the kind hospitality of their Groups, wherein part of this work
was performed.

\newpage
% =========== big frame, title etc. =======
\setlength{\unitlength}{0.1mm}
\begin{picture}(1600,1500)
\put(0,0){\framebox(1600,1500){ }}
% =========== small frame, labeled axis ===
\put(300,250){\begin{picture}( 1200,1200)
\put(0,0){\framebox( 1200,1200){ }}
% =========== x and y axis ================
% .......SAXIX........
%  JY=    2
\multiput(  300.00,0)(  300.00,0){   4}{\line(0,1){25}}
\multiput(    0.00,0)(   30.00,0){  41}{\line(0,1){10}}
\multiput(  300.00,1200)(  300.00,0){   4}{\line(0,-1){25}}
\multiput(    0.00,1200)(   30.00,0){  41}{\line(0,-1){10}}
\put( 300,-25){\makebox(0,0)[t]{\small $    5.000 $}}
\put( 600,-25){\makebox(0,0)[t]{\small $   10.000 $}}
\put( 900,-25){\makebox(0,0)[t]{\small $   15.000 $}}
\put(1200,-25){\makebox(0,0)[t]{\small $   20.000 $}}
% .......SAXIY........
%  JY=    5
\multiput(0,    0.00)(0,  240.00){   6}{\line(1,0){25}}
\multiput(0,   24.00)(0,   24.00){  50}{\line(1,0){10}}
\multiput(1200,    0.00)(0,  240.00){   6}{\line(-1,0){25}}
\multiput(1200,   24.00)(0,   24.00){  50}{\line(-1,0){10}}
\put(-25,   0){\makebox(0,0)[r]{\small $    0.000\cdot 10^{   5} $}}
\put(-25, 240){\makebox(0,0)[r]{\small $    1.000\cdot 10^{   5} $}}
\put(-25, 480){\makebox(0,0)[r]{\small $    2.000\cdot 10^{   5} $}}
\put(-25, 720){\makebox(0,0)[r]{\small $    3.000\cdot 10^{   5} $}}
\put(-25, 960){\makebox(0,0)[r]{\small $    4.000\cdot 10^{   5} $}}
\put(-25,1200){\makebox(0,0)[r]{\small $    5.000\cdot 10^{   5} $}}
\end{picture}}% end of plotting labeled axis
%========== next plot (line) ==========
%==== HISTOGRAM ID=300001
%  first
\put(300,250){\begin{picture}( 1200,1200)
% ========== plotting primitives ==========
\thinlines
\newcommand{\x}[3]{\put(#1,#2){\line(1,0){#3}}}
\newcommand{\y}[3]{\put(#1,#2){\line(0,1){#3}}}
\newcommand{\z}[3]{\put(#1,#2){\line(0,-1){#3}}}
\newcommand{\e}[3]{\put(#1,#2){\line(0,1){#3}}}
\y{   0}{   0}{1021}\x{   0}{1021}{  59}
\z{  59}{1021}{ 146}\x{  59}{ 875}{  60}
\z{ 119}{ 875}{ 503}\x{ 119}{ 372}{  60}
\z{ 179}{ 372}{ 268}\x{ 179}{ 104}{  60}
\z{ 239}{ 104}{  83}\x{ 239}{  21}{  60}
\z{ 299}{  21}{  21}\x{ 299}{   0}{  60}
\y{ 359}{   0}{   0}\x{ 359}{   0}{  60}
\y{ 419}{   0}{   0}\x{ 419}{   0}{  60}
\y{ 479}{   0}{   0}\x{ 479}{   0}{  60}
\y{ 539}{   0}{   0}\x{ 539}{   0}{  60}
\y{ 599}{   0}{   0}\x{ 599}{   0}{  60}
\y{ 659}{   0}{   0}\x{ 659}{   0}{  60}
\y{ 719}{   0}{   0}\x{ 719}{   0}{  60}
\y{ 779}{   0}{   0}\x{ 779}{   0}{  60}
\y{ 839}{   0}{   0}\x{ 839}{   0}{  60}
\y{ 899}{   0}{   0}\x{ 899}{   0}{  60}
\y{ 959}{   0}{   0}\x{ 959}{   0}{  60}
\y{1019}{   0}{   0}\x{1019}{   0}{  60}
\y{1079}{   0}{   0}\x{1079}{   0}{  60}
\y{1139}{   0}{   0}\x{1139}{   0}{  60}
\end{picture}} % end of plotting histogram
\end{picture} % close entire picture
\begin{center}
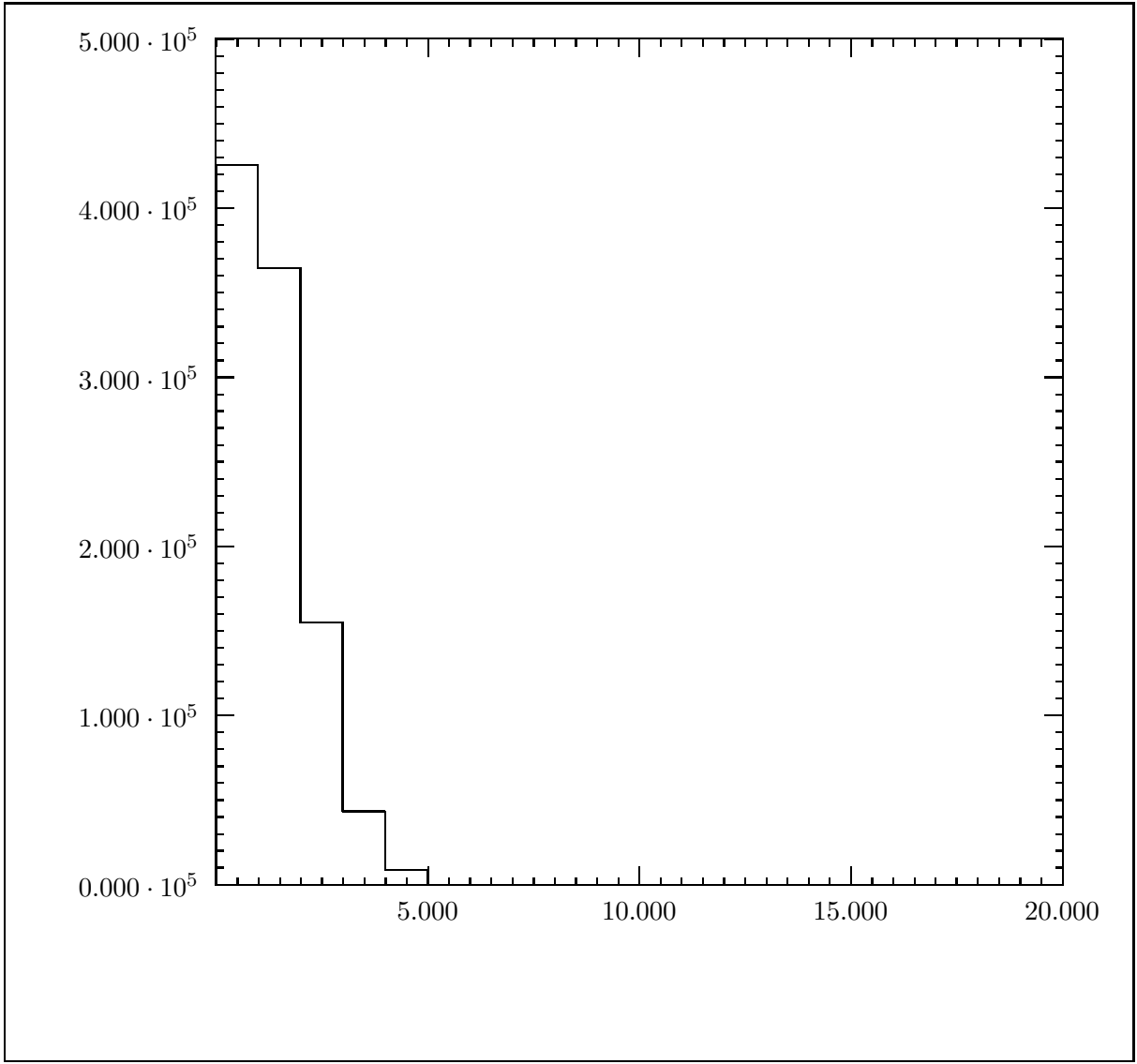
Fig.1: Histogram of the photon multiplicity in $uu\to uu+
n\gamma$ for \etatwet, \sqstev.
\end{center}
\newpage
% =========== big frame, title etc. =======
\setlength{\unitlength}{0.1mm}
\begin{picture}(1600,1500)
\put(0,0){\framebox(1600,1500){ }}
% =========== small frame, labeled axis ===
\put(300,250){\begin{picture}( 1200,1200)
\put(0,0){\framebox( 1200,1200){ }}
% =========== x and y axis ================
% .......SAXIX........
%  JY=    1
\multiput(  300.00,0)(  300.00,0){   4}{\line(0,1){25}}
\multiput(    0.00,0)(   30.00,0){  41}{\line(0,1){10}}
\multiput(  300.00,1200)(  300.00,0){   4}{\line(0,-1){25}}
\multiput(    0.00,1200)(   30.00,0){  41}{\line(0,-1){10}}
\put( 300,-25){\makebox(0,0)[t]{\small $    2.500\cdot 10^{  -1} $}}
\put( 600,-25){\makebox(0,0)[t]{\small $    5.000\cdot 10^{  -1} $}}
\put( 900,-25){\makebox(0,0)[t]{\small $    7.500\cdot 10^{  -1} $}}
\put(1200,-25){\makebox(0,0)[t]{\small $   10.000\cdot 10^{  -1} $}}
% .......SAXIY........
%  JY=    9
\multiput(0,    0.00)(0,  266.67){   5}{\line(1,0){25}}
\multiput(0,   26.67)(0,   26.67){  45}{\line(1,0){10}}
\multiput(1200,    0.00)(0,  266.67){   5}{\line(-1,0){25}}
\multiput(1200,   26.67)(0,   26.67){  45}{\line(-1,0){10}}
\put(-25,   0){\makebox(0,0)[r]{\small $    0.000\cdot 10^{   5} $}}
\put(-25, 266){\makebox(0,0)[r]{\small $    2.000\cdot 10^{   5} $}}
\put(-25, 533){\makebox(0,0)[r]{\small $    4.000\cdot 10^{   5} $}}
\put(-25, 800){\makebox(0,0)[r]{\small $    6.000\cdot 10^{   5} $}}
\put(-25,1066){\makebox(0,0)[r]{\small $    8.000\cdot 10^{   5} $}}
\end{picture}}% end of plotting labeled axis
%========== next plot (line) ==========
%==== HISTOGRAM ID=300005
%  first
\put(300,250){\begin{picture}( 1200,1200)
% ========== plotting primitives ==========
\thinlines
\newcommand{\x}[3]{\put(#1,#2){\line(1,0){#3}}}
\newcommand{\y}[3]{\put(#1,#2){\line(0,1){#3}}}
\newcommand{\z}[3]{\put(#1,#2){\line(0,-1){#3}}}
\newcommand{\e}[3]{\put(#1,#2){\line(0,1){#3}}}
\y{   0}{   0}{1095}\x{   0}{1095}{  39}
\z{  39}{1095}{1046}\x{  39}{  49}{  40}
\z{  79}{  49}{  20}\x{  79}{  29}{  40}
\z{ 119}{  29}{   7}\x{ 119}{  22}{  40}
\z{ 159}{  22}{   4}\x{ 159}{  18}{  40}
\z{ 199}{  18}{   3}\x{ 199}{  15}{  40}
\z{ 239}{  15}{   2}\x{ 239}{  13}{  40}
\z{ 279}{  13}{   1}\x{ 279}{  12}{  40}
\z{ 319}{  12}{   1}\x{ 319}{  11}{  40}
\y{ 359}{  11}{   0}\x{ 359}{  11}{  40}
\z{ 399}{  11}{   1}\x{ 399}{  10}{  40}
\y{ 439}{  10}{   0}\x{ 439}{  10}{  40}
\y{ 479}{  10}{   0}\x{ 479}{  10}{  40}
\y{ 519}{  10}{   0}\x{ 519}{  10}{  40}
\y{ 559}{  10}{   1}\x{ 559}{  11}{  40}
\z{ 599}{  11}{  11}\x{ 599}{   0}{  40}
\y{ 639}{   0}{   0}\x{ 639}{   0}{  40}
\y{ 679}{   0}{   0}\x{ 679}{   0}{  40}
\y{ 719}{   0}{   0}\x{ 719}{   0}{  40}
\y{ 759}{   0}{   0}\x{ 759}{   0}{  40}
\y{ 799}{   0}{   0}\x{ 799}{   0}{  40}
\y{ 839}{   0}{   0}\x{ 839}{   0}{  40}
\y{ 879}{   0}{   0}\x{ 879}{   0}{  40}
\y{ 919}{   0}{   0}\x{ 919}{   0}{  40}
\y{ 959}{   0}{   0}\x{ 959}{   0}{  40}
\y{ 999}{   0}{   0}\x{ 999}{   0}{  40}
\y{1039}{   0}{   0}\x{1039}{   0}{  40}
\y{1079}{   0}{   0}\x{1079}{   0}{  40}
\y{1119}{   0}{   0}\x{1119}{   0}{  40}
\y{1159}{   0}{   0}\x{1159}{   0}{  40}
\end{picture}} % end of plotting histogram
\end{picture} % close entire picture
\begin{center}
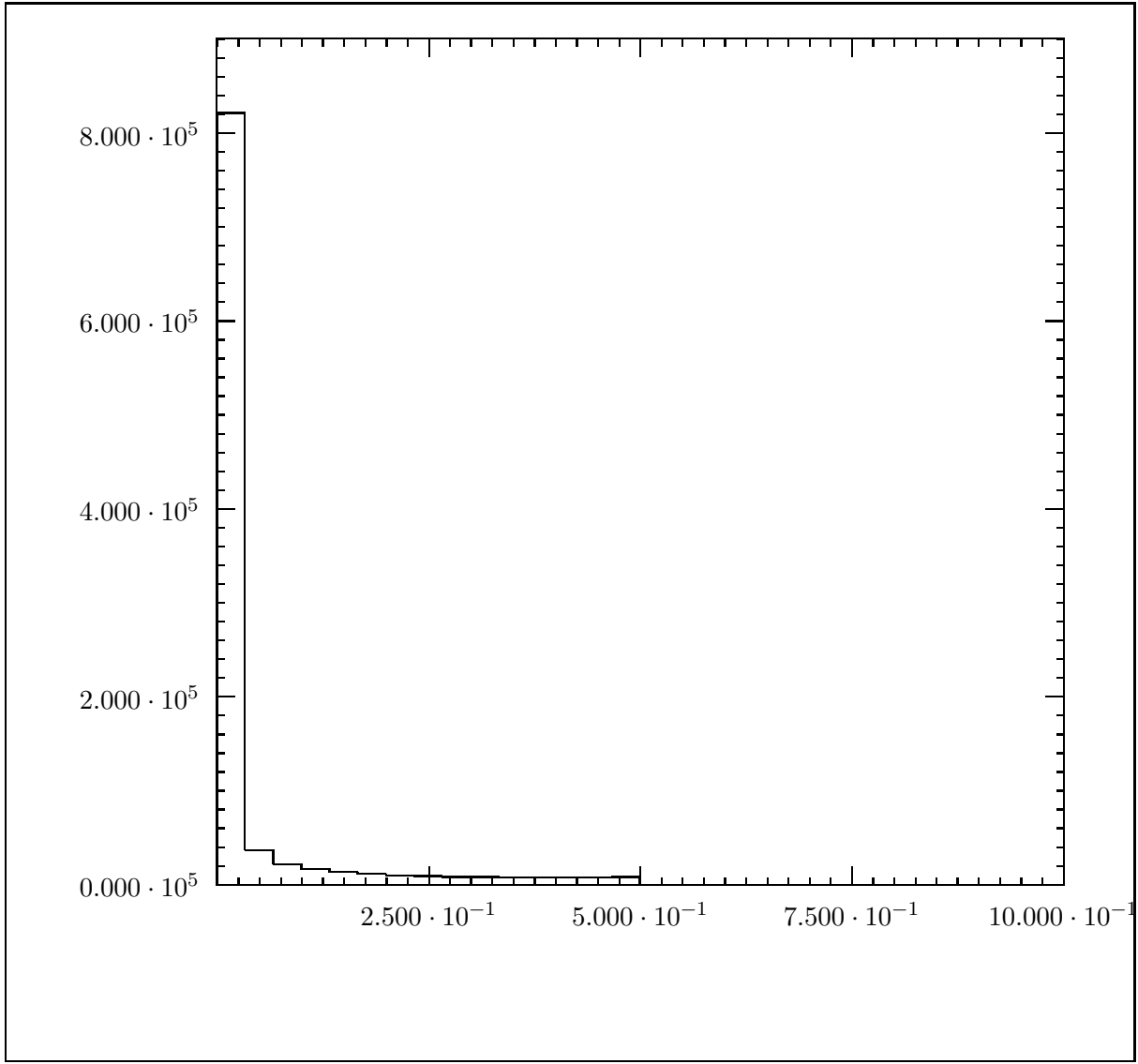
Fig.2: $v$-distribution for $uu\to uu+n\gamma$,
where $v=(s-s')/s$
and $s'=(q_1+q_2)^2$ is the squared final $uu$ invariant mass.
Here, \etatwet, \sqstev.
\end{center}
\newpage
% =========== big frame, title etc. =======
\setlength{\unitlength}{0.1mm}
\begin{picture}(1600,1500)
\put(0,0){\framebox(1600,1500){ }}
% =========== small frame, labeled axis ===
\put(300,250){\begin{picture}( 1200,1200)
\put(0,0){\framebox( 1200,1200){ }}
% =========== x and y axis ================
% .......SAXIX........
%  JY=    1
\multiput(  300.00,0)(  300.00,0){   4}{\line(0,1){25}}
\multiput(    0.00,0)(   30.00,0){  41}{\line(0,1){10}}
\multiput(  300.00,1200)(  300.00,0){   4}{\line(0,-1){25}}
\multiput(    0.00,1200)(   30.00,0){  41}{\line(0,-1){10}}
\put( 300,-25){\makebox(0,0)[t]{\small $    2.500\cdot 10^{  -1} $}}
\put( 600,-25){\makebox(0,0)[t]{\small $    5.000\cdot 10^{  -1} $}}
\put( 900,-25){\makebox(0,0)[t]{\small $    7.500\cdot 10^{  -1} $}}
\put(1200,-25){\makebox(0,0)[t]{\small $   10.000\cdot 10^{  -1} $}}
% .......SAXIY........
%  JY=    6
\multiput(0,    0.00)(0,  200.00){   7}{\line(1,0){25}}
\multiput(0,   20.00)(0,   20.00){  60}{\line(1,0){10}}
\multiput(1200,    0.00)(0,  200.00){   7}{\line(-1,0){25}}
\multiput(1200,   20.00)(0,   20.00){  60}{\line(-1,0){10}}
\put(-25,   0){\makebox(0,0)[r]{\small $    0.000\cdot 10^{   5} $}}
\put(-25, 200){\makebox(0,0)[r]{\small $    1.000\cdot 10^{   5} $}}
\put(-25, 400){\makebox(0,0)[r]{\small $    2.000\cdot 10^{   5} $}}
\put(-25, 600){\makebox(0,0)[r]{\small $    3.000\cdot 10^{   5} $}}
\put(-25, 800){\makebox(0,0)[r]{\small $    4.000\cdot 10^{   5} $}}
\put(-25,1000){\makebox(0,0)[r]{\small $    5.000\cdot 10^{   5} $}}
\put(-25,1200){\makebox(0,0)[r]{\small $    6.000\cdot 10^{   5} $}}
\end{picture}}% end of plotting labeled axis
%========== next plot (line) ==========
%==== HISTOGRAM ID=300003
%  first
\put(300,250){\begin{picture}( 1200,1200)
% ========== plotting primitives ==========
\thinlines
\newcommand{\x}[3]{\put(#1,#2){\line(1,0){#3}}}
\newcommand{\y}[3]{\put(#1,#2){\line(0,1){#3}}}
\newcommand{\z}[3]{\put(#1,#2){\line(0,-1){#3}}}
\newcommand{\e}[3]{\put(#1,#2){\line(0,1){#3}}}
\y{   0}{   0}{1119}\x{   0}{1119}{  39}
\z{  39}{1119}{1105}\x{  39}{  14}{  40}
\z{  79}{  14}{   8}\x{  79}{   6}{  40}
\z{ 119}{   6}{   2}\x{ 119}{   4}{  40}
\z{ 159}{   4}{   2}\x{ 159}{   2}{  40}
\z{ 199}{   2}{   1}\x{ 199}{   1}{  40}
\z{ 239}{   1}{   1}\x{ 239}{   0}{  40}
\y{ 279}{   0}{   0}\x{ 279}{   0}{  40}
\y{ 319}{   0}{   0}\x{ 319}{   0}{  40}
\y{ 359}{   0}{   0}\x{ 359}{   0}{  40}
\y{ 399}{   0}{   0}\x{ 399}{   0}{  40}
\y{ 439}{   0}{   0}\x{ 439}{   0}{  40}
\y{ 479}{   0}{   0}\x{ 479}{   0}{  40}
\y{ 519}{   0}{   0}\x{ 519}{   0}{  40}
\y{ 559}{   0}{   0}\x{ 559}{   0}{  40}
\y{ 599}{   0}{   0}\x{ 599}{   0}{  40}
\y{ 639}{   0}{   0}\x{ 639}{   0}{  40}
\y{ 679}{   0}{   0}\x{ 679}{   0}{  40}
\y{ 719}{   0}{   0}\x{ 719}{   0}{  40}
\y{ 759}{   0}{   0}\x{ 759}{   0}{  40}
\y{ 799}{   0}{   0}\x{ 799}{   0}{  40}
\y{ 839}{   0}{   0}\x{ 839}{   0}{  40}
\y{ 879}{   0}{   0}\x{ 879}{   0}{  40}
\y{ 919}{   0}{   0}\x{ 919}{   0}{  40}
\y{ 959}{   0}{   0}\x{ 959}{   0}{  40}
\y{ 999}{   0}{   0}\x{ 999}{   0}{  40}
\y{1039}{   0}{   0}\x{1039}{   0}{  40}
\y{1079}{   0}{   0}\x{1079}{   0}{  40}
\y{1119}{   0}{   0}\x{1119}{   0}{  40}
\y{1159}{   0}{   0}\x{1159}{   0}{  40}
\end{picture}} % end of plotting histogram
\end{picture} % close entire picture
\begin{center}
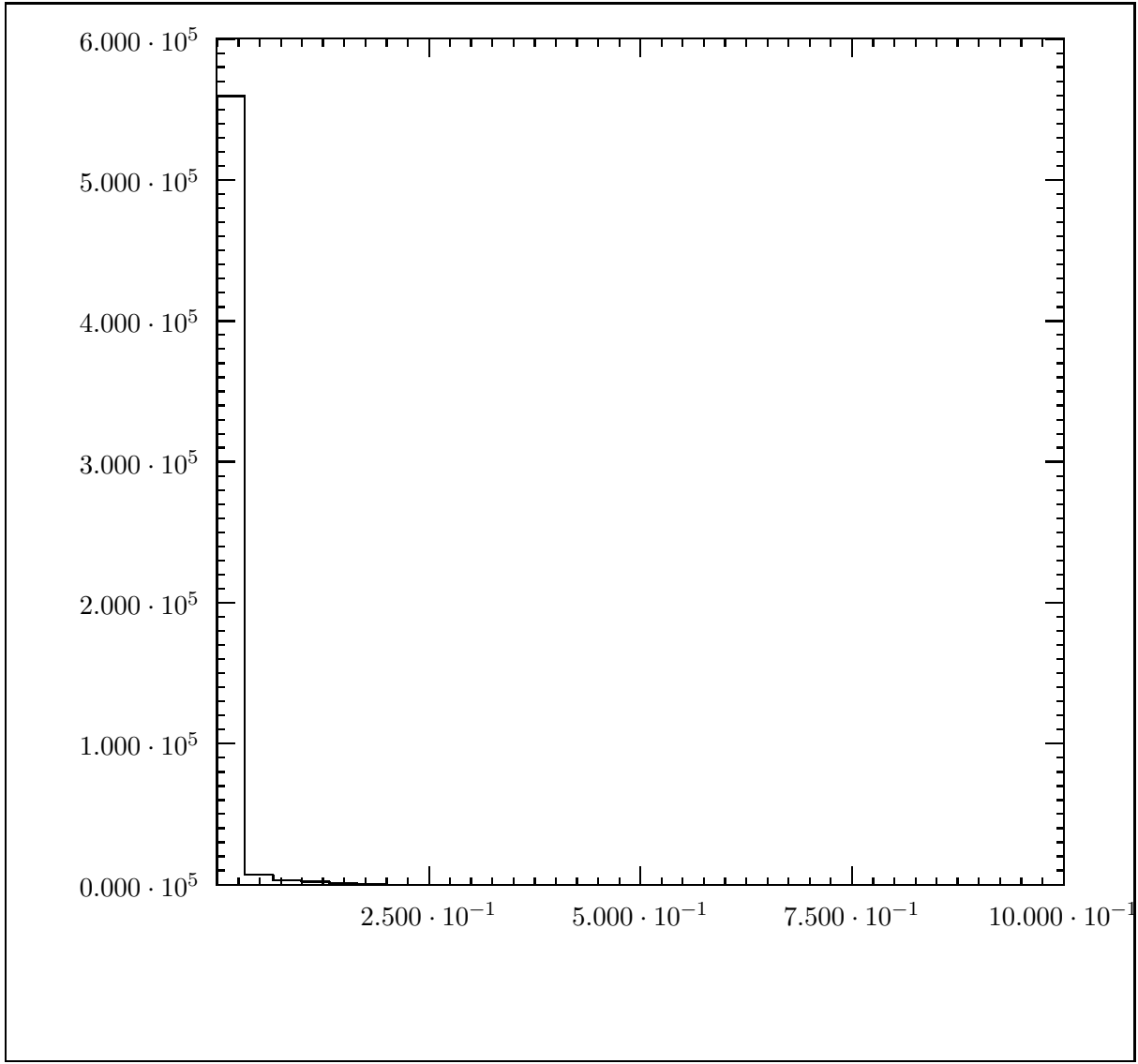
Fig.3: Total squared transverse momentum distribution
of the photons in
$uu\to uu+n\gamma$ for \etatwet\ in units of $s$, \sqstev.
\end{center}

\newpage
\begin{thebibliography}{99}

\bibitem{cjw} See, e.g., S.~Jadach and B.~F.~L. Ward, \cpc{56} (1990) 351;
\bibitem{yfs} D.~R. Yennie, S.~C. Frautschi, and H.~Suura, \anp{13} (1961) 379;
K.~T. Mahanthappa, \prz{126} (1962) 329.
\pr{40} (1989) 3852; {\it ibid.} {\bf 38} (1988) 2897, and references therein.
\bibitem{djs} D.~B. DeLaney, \etal, preprint UTHEP-92-0101, 1992.
\bibitem{ddl} D.~B. DeLaney, \etal, to appear.
\bibitem{sjb} S.~Jadach, \etal, \pr{42} (1990) 2977.
\bibitem{jwb} S.~Jadach and B.~F.~L. Ward, \pr{40} (1989) 3582.
\bibitem{eei} See, e.g., E.~Eichten, \etal, \rmp{56} (1991) 579.
\bibitem{fep} See, e.g., F.~E. Paige and D.~Protopopescu, in {\sl Snowmass
Summer Study 1986}, ed. R.~Donaldson and J.~Marx (American Physical Society,
New York 1988) p.320; H.-U.~Bengtsson and T.~Sjostrand, \cpc{46} (1987) 43.
\end{thebibliography}
\end{document}